\documentclass[aps,prb,twocolumn]{revtex4-2}

\usepackage{amsmath}
\usepackage{graphicx}
\usepackage{dcolumn}
\usepackage{bm}
\usepackage{mathrsfs}
\usepackage{mhchem}
\usepackage{subfigure}
\usepackage{color}
\usepackage{booktabs}
\usepackage{epstopdf}
\usepackage{multirow}

\usepackage{chemarrow}
\usepackage{extarrows}
\usepackage{mathtools}
\usepackage{natbib}

\begin{document}

\title{Control of active polymeric filaments by chemically-powered nanomotors}

\author{Liyan Qiao}
 \email{qiaoliyan@hdu.edu.cn}
\affiliation{Department of Physics, Hangzhou Dianzi University, Hangzhou 310018, China}

\author{Raymond Kapral}
 \email{r.kapral@utoronto.ca}
 \affiliation{Chemical Physics Theory Group, Department of Chemistry, University of Toronto, Toronto, Ontario M5S 3H6, Canada}

\date{\today}

\begin{abstract}
Active materials with distinctive nonequilibrium properties have diverse materials science applications. Active systems are common in living matter, such as the filament network in the cell that is activated by molecular motors, and in materials science as exemplified by hydrogels activated by chemical reactions. Here we describe another class of active polymeric filament systems where the filaments are activated by embedded chemically-powered nanomotors that have catalytic and noncatalytic parts. Chemical reactions on the catalytic surfaces produce forces that act on the polymeric filaments. By changing the nonequilibrium conditions these forces can be made to change sign and thereby compress or expand the filaments. The embedded motors provide both the source of activity and the means to control the filament conformational structure. As an example of control, we show that oscillatory variations of the chemical constraints yield gel-like networks that oscillate between expanded or compressed forms, much like those of hydrogels.
\end{abstract}


\maketitle
\section{Introduction}

Networks and solutions of semiflexible polymers arise in a variety of contexts and have a wide range of materials science and biological applications.  In the biological realm, active cytoskeletal networks play important roles in cell function, such as cellular transport and organization~\cite{alberts-cell2002,howard2001,2006A,0Cell}.  Filamentous actin, microtubules, and other protein filaments make up the cytoskeletal network, which, activated by the motions of out-of-equilibrium molecular motors, is responsible for many of the mechanical functions of cells~\cite{2014Modeling}. The unusual material properties of such active biopolymer networks have stimulated the search for new synthetic active materials.

The design of active functional materials and systems capable of performing specific tasks in response to internal and external signals is an important objective of research in this area~\cite{2018Intelligent}. Synthetic polymer gels have been used to construct such active systems~\cite{2014Evol,2017Oscillatory,2018Pulsatile}, and smart polymeric materials that exhibit biomimetic behavior have been made. The chemomechanical coupling between nonlinear oscillating chemical reactions and the mechanical properties of gels has been exploited to construct self-oscillating gels that undergo spontaneous, homogeneous, periodic swelling and de-swelling in a closed solution under constant conditions without the need of external stimuli~\cite{1998Self,2000In}. The mechanism that gives rise to the chemomechanical self-oscillation in hydrogels activated by the Belousov-Zhabotinsky reaction involves changes in the gel structure induced by the periodic redox changes in the oxidized and reduced states of the bound catalyst in this reaction~\cite{2003Mechanical,2006Modeling} These gels have been proposed as analogs of nerve pulses, the rhythmic beating of cardiac cells, and deformable muscles in animals~\cite{Lin2016Retrograde}.

Active biological filament networks derive their activity from molecular motors that attach and detach from the biofilaments. Likewise, synthetic active motors can attach to filaments in a network and such attachment not only tames the detrimental effects of orientational Brownian motion but also changes the properties of the network~\cite{2020Active}. By contrast, here we consider active filament systems where the constituent filaments themselves possess active properties because they contain embedded synthetic nanomotors. Filaments with active elements have been made in the laboratory by joining chemically synthesized small colloidal or Janus particles~\cite{ramirez2013polloidal,biswas2017linking,V2017,nishiguchi2018flagellar}. Theoretical investigations of freely-moving active filaments~\cite{ghosh2014dynamics,isele2015self,winkler2017active}, active polymers~\cite{bianco2018globulelike,foglino2019non,locatelli2021activity}, clamped beating filaments with spontaneous oscillations~\cite{laskar2013hydrodynamic,sarkar2017spontaneous} and the collective behavior of active worm-like chain filaments~\cite{2018Collective} have been carried out. In these systems, the active driving process and polymer conformational state play important roles~\cite{2020the}. The coupling of thermal and active noise, hydrodynamic interactions and polymer conformational changes suggests that interesting structural and dynamical features may arise in networks of such active filaments.

The active filaments we consider are constructed by inserting chemically-powered nanomotors that move by a diffusiophoretic mechanism into a semiflexible polymer chain. Through the diffusiophoretic mechanism, catalytic chemical reactions on the motor produce diffusiophoretic forces that act on the filament segments giving rise to chemomechanical coupling that can alter the conformational state of the filament. We show that these forces can be changed by chemical constraints, allowing control of the conformational structure of the polymeric filament.

Section~\ref{sec:single filament} describes how the active filaments are constructed, the diffusiophoretic mechanism, and how the conformational dynamics of a single active filament responds to constraints that change the diffusiophoretic forces the embedded motors exert on the filament. Section~\ref{sec:network} considers systems of many active filaments with embedded motors, and it is shown that the conformational system states are qualitatively different when the embedded motors tend to elongate or contract the constituent filaments. The response of many-filament systems to periodic variations in the concentration constraints is the topic of Sec.~\ref{sec:tuning} where oscillating gel-like dynamics is observed. The conclusions are given in Sec.~\ref{sec:conclusion}, followed by an Appendix where additional details of the model construction and simulation algorithm are given.

\section{Conformational dynamics of active polymeric filaments}\label{sec:single filament}

\begin{figure}[htbp]
\centering
\vspace{-0.3cm}
\resizebox{0.7\columnwidth}{!}{%
\includegraphics{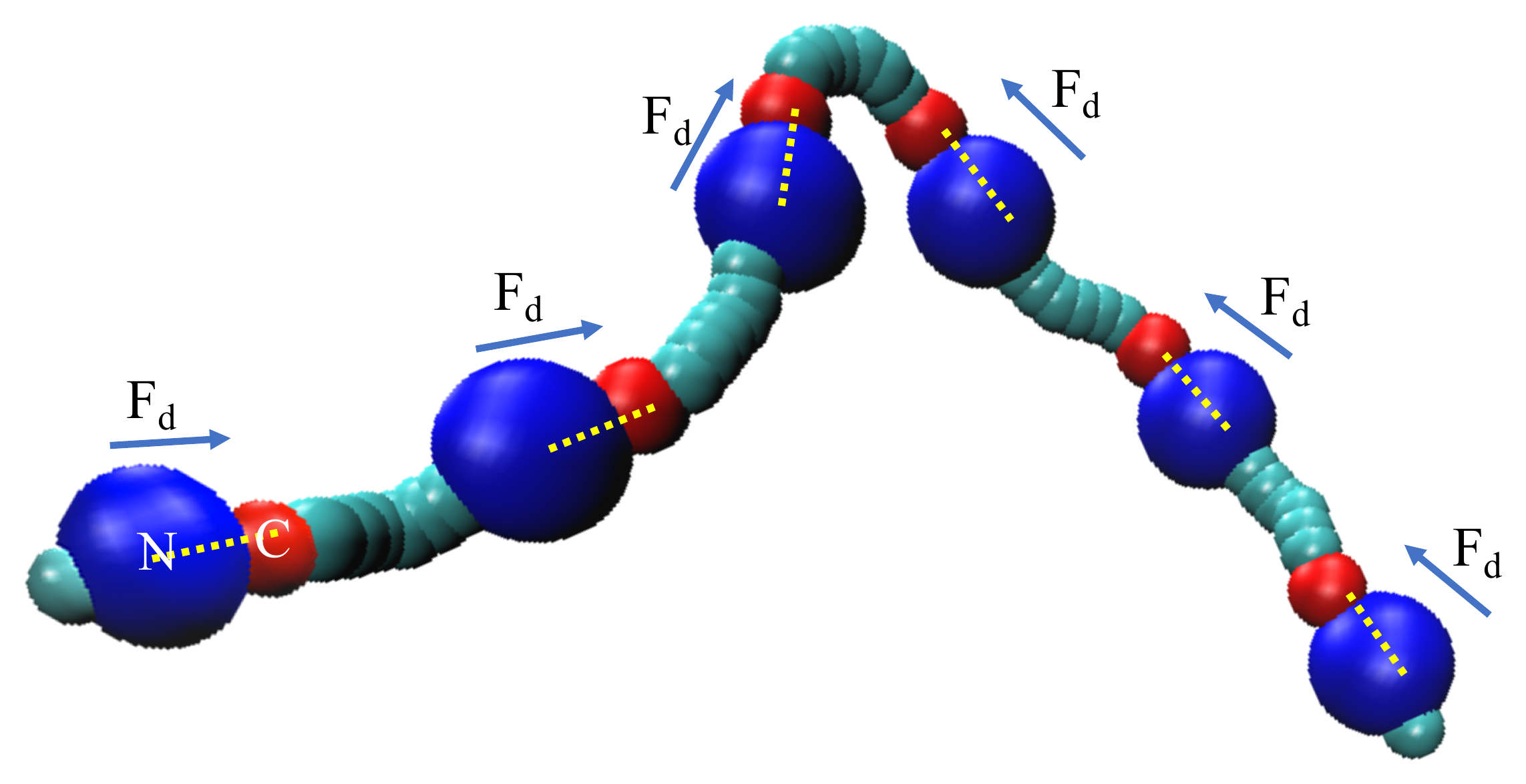} }
\vspace{-0.5cm}
\caption{\label{fig:fm-bm_6seg} A filament with six chemically-powered dimer motor segments.  The left and right three motors, respectively, are oriented in directions opposite to each other. Each motor segment consists of catalytic (red) and noncatalytic (blue) beads. The arrows indicate the directions in which the forces act. In this and all subsequent figures the data is reported in the dimensionless units based on energy in units of $\epsilon$, mass in units of $m$ and distance in units of $\sigma$.}
\end{figure}

In this section we describe how the active filaments are constructed and characterize their properties. We adopt a coarse-grained model where heteropolymeric filaments are built from two basic building blocks: $F$ monomers (beads) that are connected to form homopolymeric segments, and dimer motors that serve as links between the homopolymeric segments or as end groups. The dimer motors are themselves constructed from linked catalytic $C$ and noncatalytic $N$ beads.~\cite{Ruckner2007} All beads in the heterofilament are connected by stiff harmonic springs but the spring constants for $FF$ links are weaker than those for $FN$, $FC$ and $CN$ bonds. Three-body potentials with bending energy characterized by $\kappa_b$ determine the stiffness of the filaments. In addition, there are pair-wise, short-range, repulsive interactions among all beads to insure that the chains are self-avoiding. Figure~\ref{fig:fm-bm_6seg} shows an example of a filament with a total of $N_f=44$ beads constructed from homopolymer segments and six dimer motors.

The filaments are in a solution of $A$ and $B$ species that interact with the filament beads through short-range repulsive intermolecular potentials. The solvent species interact among themselves through multiparticle collisions~\cite{Malevanets_Kapral_99,kapral:08,gompper:2009}, and the evolution of the entire system is carried out by combining molecular dynamics and multiparticle collision dynamics~\cite{Malevanets_Kapral_00}. Since the dynamics conserves momentum (and mass and energy) hydrodynamic interactions among the polymer beads are taken into account. Full details of the intermolecular potentials and simulation algorithm are given in the Appendix.

The dimer motors are propelled in solution by a diffusiophoretic mechanism~\cite{Anderson_89,Golestanian2007,OPD17,debuyl:13,kapral2013perspective,GK19} where chemical reactions on the catalytic sphere produce local gradients of reactants and products in the vicinity of the noncatalytic sphere, which responds to these gradients to produce an active force that acts along the dimer bond~\cite{Ruckner2007,reigh2015}. While such dimer motors have been made in the laboratory from Si/Pt nanoparticles~\cite{Ozin2010}, they can be constructed from other components. For example, analogous to the half enzyme-coated silica Janus motors that have been studied experimentally~\cite{MJAHMSS15,MHPS16,ZGMS18}, dimer motors may be made by linking fully-enzyme-coated and uncoated silica nanospheres, providing motors that use a wide variety biocompatible fuels for propulsion.

For the dimer motors we consider here, we suppose that the $A$ and $B$ species interact with the noncatalytic sphere through different intermolecular potentials, and, to be consistent with microscopic reversibility~\cite{GK19},  that reversible interconversion reactions of reactant $A$ and product $B$ species, $A+C  \underset{k_-}{\stackrel{k_+}{\rightleftharpoons}} B+C$, take place on the catalytic motor bead. In this circumstance, the locally-produced asymmetric concentration gradient will give rise to a body force on the motor. Since no external forces are applied to the system, due to momentum conservation a fluid flow is generated in the vicinity of the motor that causes it to move. The motor propulsion velocity $\bm{V}_d$ is proportional to the surface average over the motor of the concentration gradients,
\begin{equation}\label{eq:Vd}
\bm{V}_d  \propto  \sum_{k=A}^B {\mathcal C}_k \overline{\bm{\nabla_s} c_k(\bm{r})}^S.
\end{equation}
 The specific forms for the prefactors ${\mathcal C}_k$ in the velocity expression have been computed analytically for both Janus~\cite{Anderson_89,Golestanian2007,OPD17,GK18} and sphere dimer~\cite{popescu2011,reigh2015,michelin2015} motors. The diffusiophoretic velocity is related to the diffusiophoretic force by $\bm{V}_d =\bm{F}_d/\zeta$, where $\zeta$ is the friction coefficient of the motor in solution.~\cite{GK18,GK19} For the dimer motors considered here $\bm{F}_d \equiv F_d \hat{\bf u}$ is directed along the dimer bond unit vector $\hat{\bf u}$ pointing from the $N$ to $C$ beads.

When such dimer motors are components of a polymeric filament, the diffusiophoretic forces $\bm{F}_d$ that they exert on the filament will play an important role in the results we present below. In this work we choose as an example active filaments like that shown in Fig.~\ref{fig:fm-bm_6seg} where the three motors on the left and right, respectively, point with their catalytic ends in opposite directions along the filament. Consequently, the diffusiophoretic forces acting on the left and right also point in opposite directions. While our attention is restricted to filaments of this type, nanomotor segments can be inserted in the polymer chain in other ways, with different distances and orientations along the chain. In the laboratory, chains with specified dimer positions and orientations could be constructed by attaching specific chemical linker groups to the catalytic and noncatalytic dimer spheres that are designed to attach to complementary linker groups on the end $F$ groups of homopolymeric segments. We also note that in contrast to active chains built from Janus colloids where the motor orientations may be variable and difficult to control, here the dimer motor constituents are spherical, and symmetry of the dimer dictates that the diffusiophoretic force is directed along the dimer bond, which is just one of the links in the chain.

\vspace{0.1in} \noindent
{\bf Nonequilibrium conditions}: Sustained active motion is only possible if detailed balance is broken and the system is taken out of equilibrium. For the diffusiophoretic motors considered here a nonequilibrium state can be established by coupling the system to reservoirs with constant concentrations of chemical species. The reservoirs may directly control the $A$ and $B$ species concentrations, or indirectly by controlling the concentrations of other chemical species that enter the mechanisms of reactions that take place in the fluid phase. We make use of this latter method here. Specifically, we consider a reversible bimolecular reaction in the fluid with rate constants $\tilde{k}_{b\pm}$,
\begin{equation}\label{eq:fluid-reaction}
P_1 + B \underset{\tilde{k}_{b-}}{\stackrel{\tilde{k}_{b+}}{\rightleftharpoons}}  A +P_2,
\end{equation}
that involves two other chemical species, $P_{1,2}$, This reaction also interconverts $A$ and $B$ species, but by a mechanism that is different from that on the motor. The reservoirs fix the concentrations of the ``pool" species $P_{1,2}$ at constant values $\bar{c}_{P_{1,2}}$ and drive this reaction out of equilibrium. In the fluid phase, far from the motor, the mass-action chemical rate law can be written as $dc_A(t)/dt=-k_{b-}c_A(t) +k_{b+}c_B(t)$, where the effective rate coefficients $k_{b\pm}=\tilde{k}_{b\pm}\bar{c}_{P_{1,2}}$ incorporate the fixed concentrations of the pool species. Since these reservoir concentrations are under our control, we can use the effective rate coefficients $k_{b\pm}$ as control parameters to adjust the nonequilibrium state of the system. In the steady state regime the $A$ and $B$ concentrations adopt their steady state values that satisfy $-k_{b-}c_A^s +k_{b+}c_B^s=0$. Thus, we see that provided $k_+/k_- \ne k_{b+}/k_{b-}$ detailed balance will be broken and active motion will be possible.

By solving the steady state reaction-diffusion equation for the concentrations $c_A$ and $c_B$ in a system with the fluid phase reactions, $D \nabla^2 c_A -k_{b-}c_A+k_{b+}c_B=0$, subject to boundary conditions that account for  reactions on the catalytic surface, and steady state fluid concentrations, $c^s_{A,B}$, far from the motor~\cite{HSGK18}, the expression for the diffusiophoretic force that follows from Eq.~(\ref{eq:Vd}) can be expressed in terms of the steady state concentrations far from the motor to give
\begin{eqnarray}\label{eq:Fd}
{F}_d&=& f_d  (k_+{c}^s_A -k_- {c}^s_B)=f_d  k_- {c}^s_B\Big(\frac{{c}^{\rm eq}_B{c}^s_A}{{c}^{\rm eq}_A {c}^s_B}-1\Big)\nonumber\\
&=& f_d  k_- {c}^s_B(e^{A_{\rm rxn}}-1),
\end{eqnarray}
where the prefactor is chosen so that $f_d>0$, and depends on geometrical factors, the reaction-diffusion solution, and motor-fluid interaction potentials. The equilibrium condition, $k_+/k_-=c^{\rm eq}_B/c^{\rm eq}_A$ was used to write the second equality. The last equality expresses ${F}_d$ in terms of the (dimensionless) chemical affinity, $A_{\rm rxn}=-\Delta \mu/k_BT$, where the free energy of the reaction is $\Delta \mu=\mu_B-\mu_A$ with $\mu_{A,B}$ the species chemical potentials.
The diffusiophoretic force vanishes at equilibrium, while under nonequilibrium conditions its sign depends on the relative values of $k_+ {c}^s_A$ and $k_- {c}^s_B$. Although we make use of a simple fluid phase reaction here, it is possible to replace the reaction in Eq.~(\ref{eq:fluid-reaction}) by a more complex chemical network. An earlier investigation active sphere dimer motion in a medium that supports autonomous chemical oscillations generated by the Selkov enzymatic model~\cite{selkov:68,ross:81} showed how the motor and its environment interact to change the dynamics of the entire system.~\cite{Robertson2015}

This means of establishing a nonequilibrium state is used in living systems where a network of chemical reactions operating out of equilibrium due to constraints on some pool species in the network supplies the fuel that powers molecular motors in the cell. For example, the adenosine triphosphate fuel that some molecular motors use is supplied in the cell by a complex network of other enzymatic reactions operating out of equilibrium. This method is also used in laboratory experiments on the dynamics of nonlinear chemical systems that may exit in spatio-temporal states that oscillate or form chemical patterns, such as those seen in the Belousov-Zhabotinsky reaction when it is driven out-of-equilibrium by coupling to reservoirs of this type.~\cite{CFUR1988}

\vspace{0.1in} \noindent
{\bf Chemomechanical coupling}: To see how changing nonequilibrium concentration constraints can lead to chemomechanical coupling, we consider filaments with motors configured as in Fig.~\ref{fig:fm-bm_6seg}.  We adopt the convention where motors that move in a direction with their catalytic sphere at the motor head ($F_d > 0$) are termed forward-moving, while those moving with their noncatalytic sphere at the motor head ($F_d < 0$) are backward-moving. Referring to Eq.~(\ref{eq:Fd}), using the fact that ${{c}^{\rm eq}_B{c}^s_A}/{{c}^{\rm eq}_A {c}^s_B}={k_+ k_{b+}}/{k_- k_{b-} }$ and taking $k_+=k_-$ as in our simulations, if $k_{b+} > k_{b-}$ then $F_d > 0$, and the motor will move in the forward direction, and backward for $k_{b+} < k_{b-}$. Since $F_d > 0$ for forward-moving motors the diffusiophoretic forces they exert will tend to compress the filament, while backward-moving motors with $F_d < 0$ will tend to stretch the filament, giving rise to the chemomechanical coupling.

\begin{figure}[htbp]
 \centering
 \resizebox{1.0\columnwidth}{!}{
   \includegraphics[height=10.5cm,width=10.5cm]{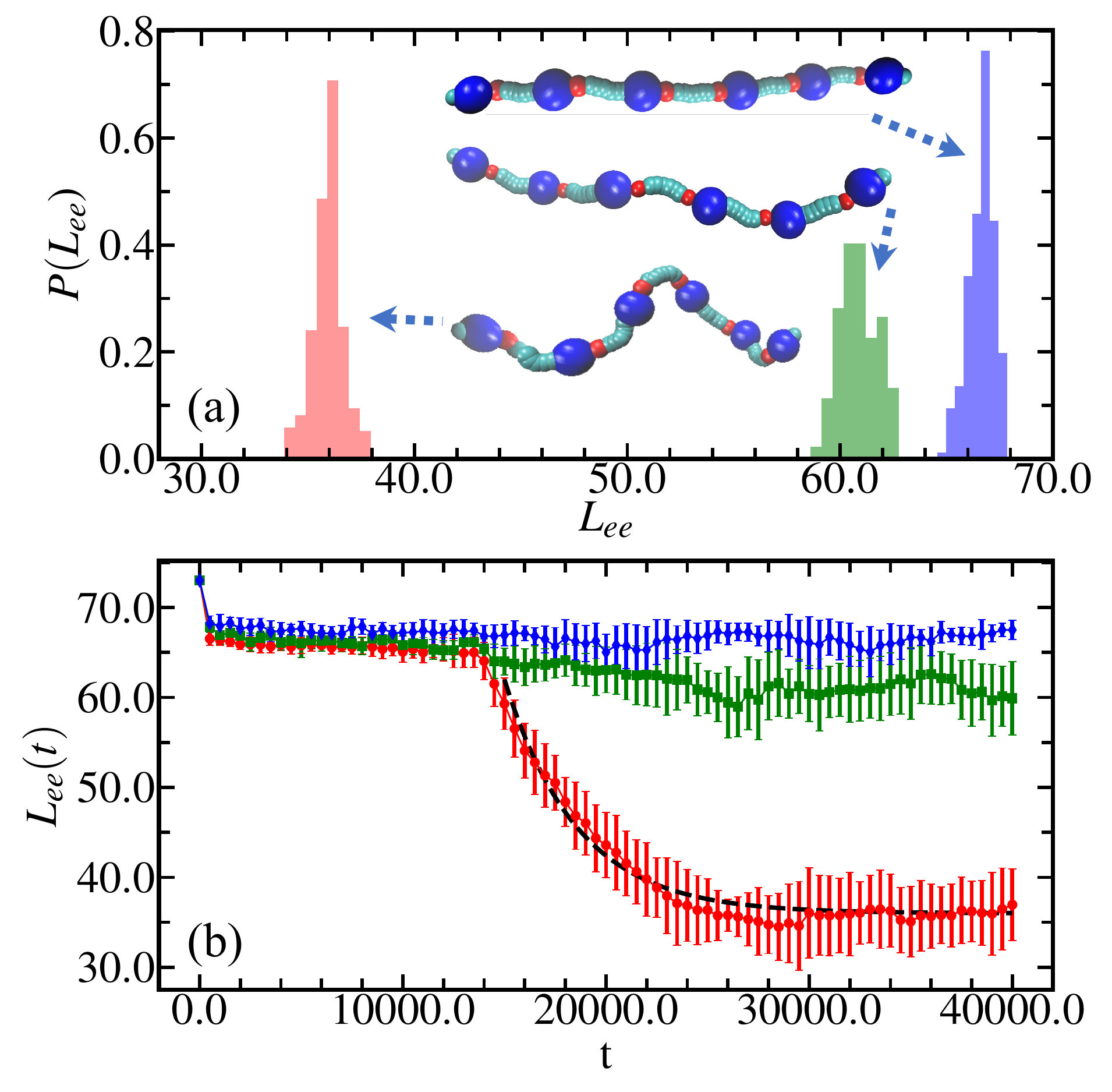}
}
  \caption{\label{fig:1fil_3figs}
(a) Plots of the probability densities $P(L_{ee})$ of the end-to-end distance for a single filament with forward-moving (red), backward-moving (blue) and inactive (green) motors. 
(b) Time evolution of the average end-to-end distance $L_{ee}(t)$ starting from a linear filament for the same cases and color coding as in (a). 
The black dashed line is a fit to the evolution of a filament with forward-moving motors using Eq.~(\ref{eq:fit}). The insets show instantaneous conformations of a single filament in the different steady state or equilibrium regimes. Results were obtained from  averages over 20 realizations the dynamics, and the error bars represent $\pm$ one standard deviation.}
\end{figure}
Figure~\ref{fig:1fil_3figs} (a) shows the probability densities $P(L_{ee})$ of the filament end-to-end length $L_{ee}$ for both forward-moving  ($(k_{b+},k_{b-})= (10^{-2},10^{-3})$) and backward-moving ($(k_{b+},k_{b-})= (10^{-3},10^{-2})$) embedded motors. This figure also compares active filaments with inactive filaments where ($(k_{b+},k_{b-})= (5.5 \times 10^{-3},5.5 \times 10^{-3})$) and the system satisfies detailed balance. One can see the distinct, strongly-localized probability distributions for the three different constraint conditions. Recall that chemical reactions still take place in the systems with an inactive filament, but they occur at chemical equilibrium. Note also that the solvated heteropolymeric chain exits in extended conformational states at equilibrium. When the activity of backward-moving motors is applied the chain elongates but the increase in chain length is small since it is difficult to increase the length of a nearly linear chain by stretching strong chemical bonds. By contrast, forward-moving motors can induce large-scale conformational changes, such as those shown in the inset in panel (a) of the figure, when starting from an extended chain conformations.

\vspace{0.1in} \noindent
{\bf Filament dynamics and structure}: Since filaments like those in Figs.~\ref{fig:fm-bm_6seg} and \ref{fig:1fil_3figs} will be used to build active networks, it is useful to further characterize their properties in solution under equilibrium and nonequilibrium conditions. Starting from an initial nonequilibrium conformation where the filament is linear chain with end-to-end length $L_{\rm init}=73$, Fig.~\ref{fig:1fil_3figs} (b) shows the evolution of the average end-to-end length, $L_{ee}(t)$, under constraints that give $F_d>0$ and $F_d<0$, as well as equilibrium conditions where $F_d=0$. For both $F_d<0$ and $F_d=0$, after a short rapid decay, the evolution leads at long times to the steady state $L_{ee}^{\rm b}=66$ and equilibrium $L_{ee}^{\rm eq}=61$ values, respectively, which are not very different from each other for the reasons described above. The figure also shows that there is transient period, $\Delta t \approx 14000$, before the differences between steady state and equilibrium state values can be distinguished. The decay for $F_d>0$ follows a similar pattern. After the rapid initial decay, the average end-to-end length remains large, $L_{ee} \approx 66$, for the same transient period, after which it decays to its final asymptotic steady state value of $L_{ee}^{\rm f}=36$.~\cite{foot:metastable} 

The evolution from an extended to a contracted filament under $F_d>0$ is approximately exponential and satisfies   \begin{equation}\label{eq:fit}
L_{ee}(t)=L_{\rm fin} +\big(L_{\rm init}-L_{\rm fin}\big)e^{-t/t_c},
\end{equation}
to a good approximation, where $L_{\rm init}$ and $L_{\rm fin}$ denote the initial and long-time values, respectively, and $t_c$ is the characteristic conformational relaxation time. We find $t_c = t^{\rm f}_c=3571$, with ($L_{\rm init}=66, L_{\rm fin}=36$). In a similar way, we may draw initial filament conformations from the system in the nonequilibrium steady state for $F_d>0$ where $L_{\rm init}=L_{ee}^{\rm f}=36$. After the constraint is changed to $F_d<0$, $L_{ee}(t)$ increases exponentially to $L_{\rm fin}=66$ with a time constant of $t_c=t^{\rm b}_c=1250$. The evolution from $L_{\rm init}=36$ under equilibrium conditions ($F_d=0$) was also computed, and lies within the error bars for that with $F_d<0$; thus, the evolution from contracted to expanded forms is aided by the intrinsic bending rigidity of the equilibrium filaments, and the application of forces that stretch the filaments play a less important role in this process. (If instead the equilibrium filaments had collapsed or partially collapsed conformations, then applied forces would be necessary to reach extended states.) 
By contrast, relaxation from extended to contracted forms takes place on a longer time scale and occurs only because the diffusiophoretic forces with $F_d>0$ act on the filament.

The dimers, even if embedded in polymer chains, still experience chemotactic and hydrodynamic interactions that lead to clustering. For such clustering to take place the entire filaments must move, and this is the origin of the filament inhomogeneity in networks with positive diffusiophoretic forces. The different nonequilibrium constraints are also reflected in translational diffusion coefficients of the center of mass of the polymer chains. The diffusion coefficients extracted from the mean square displacements of the filaments are $D_{\rm b}=7.5 \times 10^{-4}$, $D_{\rm eq}=3.0\times 10^{-4}$ and $D_{\rm f}=7.5 \times 10^{-3}$. Since the filaments for backward-moving motors are nearly linear and the motors on either side of the center oppose each other, the active contribution to diffusion will be small; still, $D_{\rm b}$ is greater  than $D_{\rm eq}$ by more than a factor of two. Filaments with forward-moving motors execute large conformational fluctuations and, in conformations with bends, the motors will produce strong net diffusiophoretic forces that contribute to active translational motion, whose direction is determined by the instantaneous conformational state of the filament. Thus, $D_{\rm f}$ is more than 20 times larger than $D_{\rm eq}$ due to this active contribution.

\begin{figure}[htbp]
  \begin{center}
\resizebox{1.0\columnwidth}{!}{
     \includegraphics[height=8.5cm,width=9.0cm]{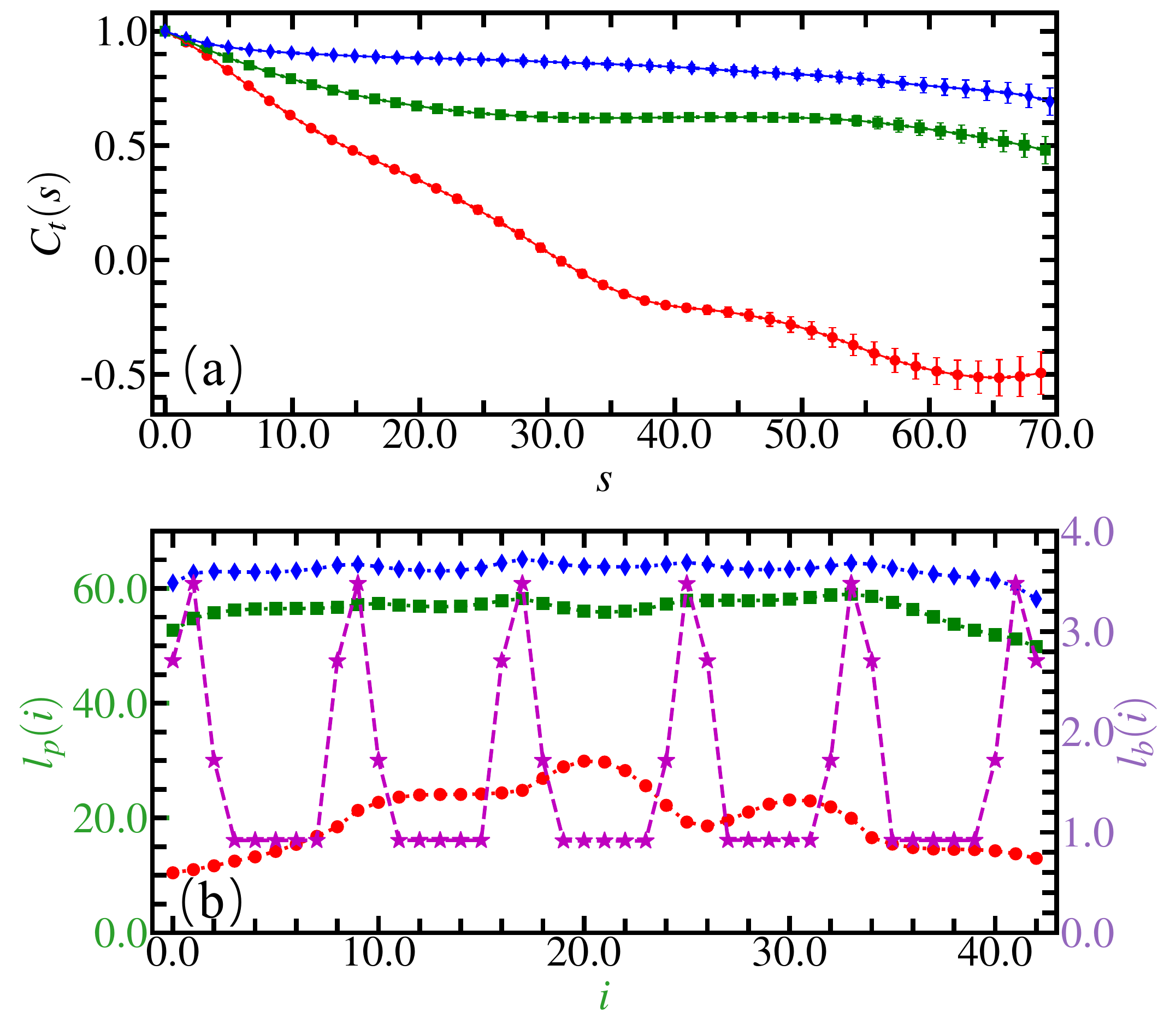}
}
         \caption{\label{fig:tt_compare_f44_f70}
 (a) Plots of the tangent correlation function $C_t(s)$ versus arc length $s$ for a single filament with forward-moving (red), backward-moving (blue) and inactive (green) motors. (b) Plots of the bond length (right ordinate, purple stars) and local persistence length (left ordinate) for the three constraint conditions as in (a) versus the chain segment index $i$.}
\end{center}
\end{figure}

The tangent correlation function,
\begin{equation}
C_t(s)=\left \langle \mathbf{t}(\tau+s)\cdot \mathbf{t}(\tau)\right \rangle,
\end{equation}
where $\mathbf{t}(s)$ is the unit tangent vector at arc length $s$ is often used to characterize the bending rigidity of polymer chains, and this function is plotted in Fig.~\ref{fig:tt_compare_f44_f70} (a) for filaments under different nonequilibrium constraints. We see that the correlations persist for inactive filaments and filaments with embedded backward-moving motors since in both cases the chains are extended and roughly linear. By contrast for filaments with embedded forward-moving motors the correlations decay much more rapidly since the chain exists in contracted, bent conformations that will cause the tangent directions to vary significantly along the chain. Such bent conformation where end motor segments have nearly opposite orientations are responsible for the negative correlations seen for large $s$. 

The persistence length $\ell_p$ provides a measure of the bending rigidity of polymers, and for ideal semi-flexible polymer chains it can be determined from the exponential decay of $C_t(s)$.~\cite{KP1949,LL1958} 
However, for short, real, polymer chains with excluded volume, like those considered in this work, $C_t(s)$ does not exhibit exponential decay, and $\ell_p$ can no longer be determined by this method.~\cite{HPB2013} Some information on the bending rigidity can be obtained by computing the local persistence length defined by
\begin{equation}
\ell_p(i) /\langle \ell_{bi}\rangle =\langle \bm{\ell}_{bi} \cdot \bm{L}_{ee} \rangle /\langle {\ell}_{bi}^2 \rangle,
\end{equation}
where $\bm{\ell}_{bi}=\bm{r}_{i+1}-\bm{r}_i$ is the bond vector of chain segment $i$ with $\bm{r}_j$ the coordinate of bead $j$ in the chain, $\bm{L}_{ee}$ is the end-to-end vector, and the average is over time and realizations of the dynamics.~\cite{HPB2013} Figure~\ref{fig:tt_compare_f44_f70} (b) plots both the bond length and local persistence length versus $i$, and shows the inhomogeneous character of both of these quantities. The expanded chains with backward-moving and inactive motors have effective persistence lengths of approximately $55-60$ that are comparable to the chain length, while that for chains with forward-moving motors is much smaller, approximately 20, and varies more strongly with $i$, reflecting the large conformational changes and flexibility of the chain interior and ends.

\section{Networks of interacting active filaments}\label{sec:network}
\begin{figure*}[htbp]
  \begin{center}
\resizebox{1.9\columnwidth}{!}{
      \includegraphics[height=6.5cm,width=17.1cm]{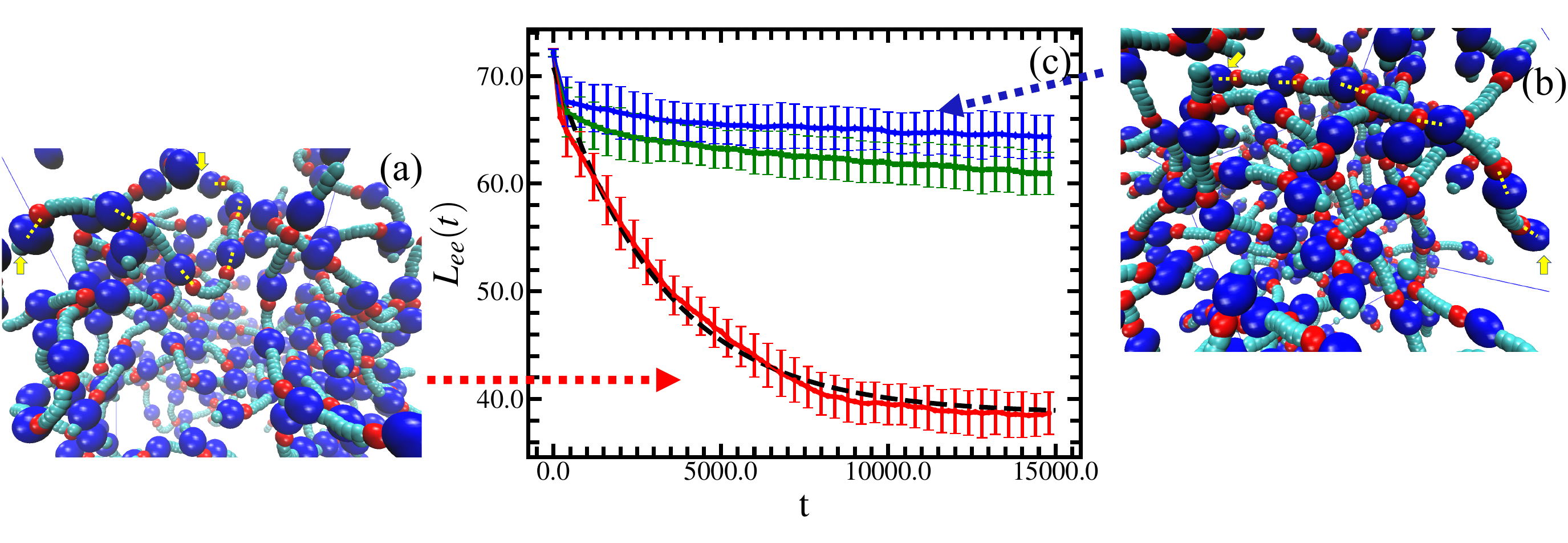}
}
         \caption{\label{fig:40f_3figs} (a) Instantaneous configuration of a system with 40 active filaments with $240$ embedded forward-moving dimer motor segments. One filament is marked by yellow arrows to show its ends, and the yellow dashed lines indicate the directions in which the filament bends. (b) The same as (a) but for active filaments with $240$ embedded backward-moving motor segments. (c) The average end-to-end distance $L_{ee}(t)$ versus $t$ for filaments in the network. The plots are for filaments with embedded forward-moving (red), backward-moving (blue) and inactive motors (green). The error bars represent $\pm$ one standard deviation computed from averages over all the filaments in the system and over ten realizations of the dynamics.}
\end{center}
\end{figure*}
We now show how the conformational structure and dynamics of many interacting active filaments depends on the nonequilibrium chemical constraints that take the system out of equilibrium. To construct the systems we study, filaments of the same length as described in the previous section are randomly distributed in the simulation box with random orientations. There are no permanent cross links between the filaments, although physical interactions can lead to transient linking. The system evolves under specified concentration constraints as discussed above. The resulting system is an entangled collection of filaments that has features similar to those of polymer gels with geometrical physical links rather than permanent links formed by chemical bonds.

A configuration extracted from the dynamics of a network of 40 active filaments with embedded forward-moving dimer motor segments is shown in Fig.~\ref{fig:40f_3figs} (a). In this image one can see the complex entangled arrangement the filaments adopt, as well its inhomogeneous structure with regions where the filaments aggregate. One filament is marked to show that its conformation is similar to that for the isolated filaments discussed in the previous section. If instead the active filaments have embedded backward-moving motor segments the chains are stretched and the corresponding interacting filament system has a different structure as shown in Fig.~\ref{fig:40f_3figs} (b). Now the system of filaments is much more homogeneous and, as can be seen in the marked filament, the individual filaments are indeed stretched. The center panel (c) shows the evolution of the average end-to-end length. The decay characteristics of $L_{ee}(t)$ are similar to those for isolated filaments discussed in the previous section, except that long transient period is no longer present, likely because of stronger fluctuations due to interactions among filaments. The evolution from extended to contracted  filament conformations of most interest is again approximately exponential, and we find the characteristic decay time, $t^{\rm f}_c = 3225$. The difference between this value and that for a single filament in bulk solution ($t^{\rm f}_c = 3571$) is not large, and the enhanced fluctuations due to the other active filaments will tend to shorten the relaxation time.(The presence of strong nonequilibrium fluctuations is presaged by the observation of enhanced diffusion of chains with active embedded motors.)

\begin{figure}[htbp]
  \begin{center}
  \resizebox{1.0\columnwidth}{!}{%
     \includegraphics[height=5.5cm,width=10.5cm]{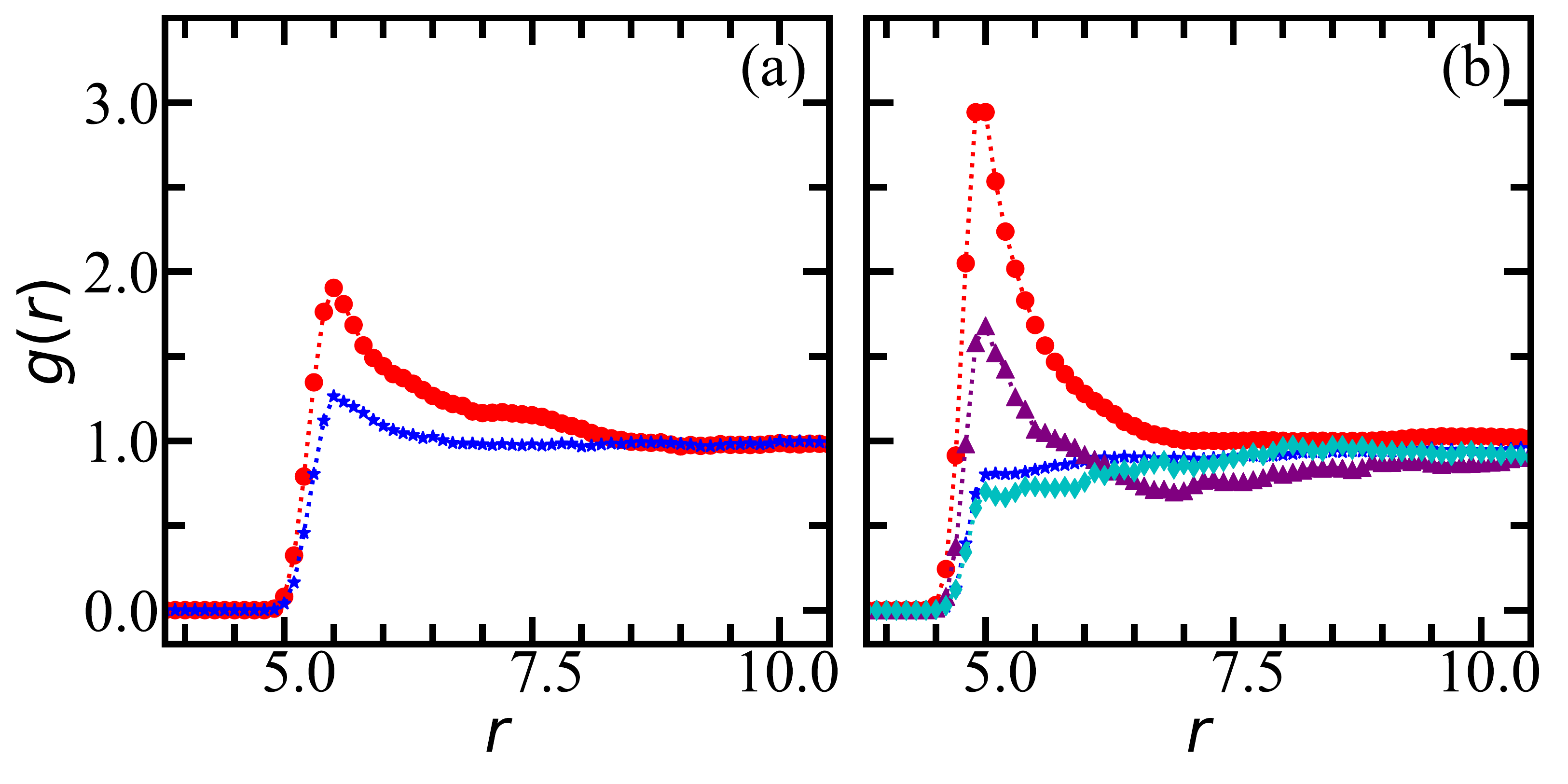}}
         \caption{\label{fig:grcu_40fila}
 The $NN$ radial distribution function $g(r)$ for (a) a collection of $240$ motors in solution, and (b) for $240$ embedded motor segments in a system with 40 filaments. The plots in these figures are for forward-moving (red circles and lines) and backward-moving (blue stars and lines) motors. Panel (b) also shows results for the  many-filament system subject to the square-wave period forcing (see Sec.~\ref{sec:tuning}) where the fluid-phase reaction rate coefficients $k_{b\pm}(t)$ oscillate with period $\tau_b=5000$. These curves were obtained averages over ten realizations of the dynamics and time averages over the time intervals $t=6000-7000$ (cyan diamonds and lines) and $t=8000-9000$ (purple triangles and lines).}
\end{center}
\end{figure}
Next, we consider the structure of the filament network in more detail. It is useful to recall that the collective dynamics of chemically-powered nanomotors in solution has been investigated extensively~\cite{Wang2015,Elgeti2015,Zoettl2016,Illien2017,CRRK18}, and an important issue that arises is the relative roles of hydrodynamic and chemotactic interactions in determining the forms that the collective behavior takes. Both of these interactions may play important roles in cluster formation, as exemplified by the observations that clustering can occur in model systems where only hydrodynamic interactions are present~\cite{saintillan2012}. By contrast, large clusters of chemically-powered Janus colloids are observed in particle-based simulations that include both hydrodynamic and chemotactic  interactions, but these clusters dissociate when the chemotactic interactions are removed.~\cite{huang2017chemotactic} In addition, it has been shown that collections of forward-moving dimer motors aggregate strongly in solution due to chemotactic attraction~\cite{Thakur2012,Qiao21,Wagner2017,Colberg2017}.

The network comprising active filaments is a many-dimer-motor system, albeit with the dimers embedded in the filaments. Nevertheless, the motors do generate a complex many-body concentration field that couples the dynamics of the motors through chemotactic interactions. In this connection it is interesting to compare the collective dynamics of $240$ dimer motors in bulk solution with that of the 40-filament system, which also has $240$ dimer motors. Figures~\ref{fig:grcu_40fila} (a) and (b) show plots of the $NN$ steady-state radial distribution function, $g(r)$,
\begin{equation}\label{eq:gnn}
g(r)= \frac{V}{4 \pi r^2 N_{\rm M}}\Big\langle \sum_{j<i=1}^{N_{\rm M}}{\delta \left ( \left | (  \mathbf{r}_{N_{i}}-\mathbf{r}_{N_{j}})  \right | -r\right )} \Big\rangle,
\end{equation}
where $r$ is the magnitude of the distance between the motor $N$ spheres, $N_{\rm M}$ is the number of motors and the angle brackets denote an average over time and realizations. In Fig.~\ref{fig:grcu_40fila} (a) for motors in bulk solution the peak in $g(r)$ for forward-moving motors indicates cluster formation, while there is a much weaker tendency for backward-moving to form clusters. Systems of $N_M=240$ dimer motors have a small volume fraction $\phi= N_M V_M/V \approx 0.045$, where the effective dimer volume is $V_M \approx 79.6$. For this small volume fraction the $g(r)$ plot indicates that forward-moving motors form small clusters with correlations extending to next-nearest  neighbors. At higher volume fractions for dimer motors with somewhat larger propulsion velocities, strong cluster formation is seen forward-moving motors, while only local small density fluctuations exist for backward-moving motors.~\cite{Colberg2017} As noted above, it has been shown that chemotactic attractive interactions that arise from the many-body concentration gradients produced from the chemical reactions on all motors are mainly responsible for cluster formation for forward-moving dimers.

Turning now to the active filament network, we see that this clustering tendency is enhanced for forward-moving motors as seen in Fig.~\ref{fig:grcu_40fila} (b), while there is no cluster formation for embedded backward-moving motors. The $g(r)$ results in this figure were obtained by counting only motors on other filaments, and excluding those on the same filament, in order to remove the built-in correlations due to those motors embedded in the same filament. The arguments given earlier for the origin of the chemomechanical coupling focused on the motors in a single filament, and relied on self-generated concentration gradients of the individual motors to produce a diffusiophoretic force that acts on the filament. The results presented in Fig.~\ref{fig:40f_3figs} show that these active forces also operate in the network of filaments. However, just as for a collection of dimer motors in bulk solution, the concentration fields of all motors contribute to the diffusiophoretic forces the filaments experience in the network. (These same many-body forces also operate in a single filament but they contribute less since the motors are separated by homopolymer chain segments in a large fraction of filament configurations.) Thus, the results suggest that the inhomogeneous structure of the network made from forward-moving motors arises from chemotactic attractive diffusiophoretic forces that tend to cause neighboring filaments to cluster. Conversely, the repulsive chemotactic interactions in filaments with backward-moving motors prevent the formation of such filament clusters.

\section{Oscillating active gel-like systems}\label{sec:tuning}

The active conformational states of the filament networks described in the previous section were shown to depend strongly on the nonequilibrium chemical constraints that give $F_d > 0$ or $F_d < 0$. The steady state concentrations $c_A^s$ and $c_B^s$ that enter
Eq.~(\ref{eq:Fd}) for $F_d$ depend on fixed concentrations of pool species $P_{1,2}$ in the reservoirs through their dependence on the effective rate coefficients, $k_{b\pm}=\tilde{k}_{b\pm}\bar{c}_{P_{1,2}}$. Here we consider situations where these reservoir concentrations vary periodically in time, consequently, so do the time-dependent effective rate coefficients,
\begin{equation}\label{eq:t-dependent-k}
k_{b\pm}(t)=\tilde{k}_{b\pm}\bar{c}_{P_{1,2}}(t).
\end{equation}

Before proceeding with the specific mechanism for fluid reactions we have adopted, it worth noting how such oscillatory reaction kinetics might be implemented in laboratory situations. For simple fluid reactions like that in Eq.~(\ref{eq:fluid-reaction}), the concentrations of the pool species can be chosen to oscillate by varying the input feeds to well-stirred reservoirs of these species. However, if more complex fluid phase reactions which support oscillatory kinetics occur in the fluid phase, then one simply needs to select the reservoir concentrations so that the reaction lies in the oscillatory regime. This is the case for the Selkov enzymatic reaction mentioned earlier~\cite{Robertson2015}, and applies to a variety of biochemical~\cite{goldbeter1996} and inorganic reaction mechanisms~\cite{epstein1998} that show oscillatory kinetics when driven out of equilibrium.

\begin{figure}[htbp]
  \begin{center}
\resizebox{1.0\columnwidth}{!}{
     \includegraphics[height=6.5cm,width=10.5cm]{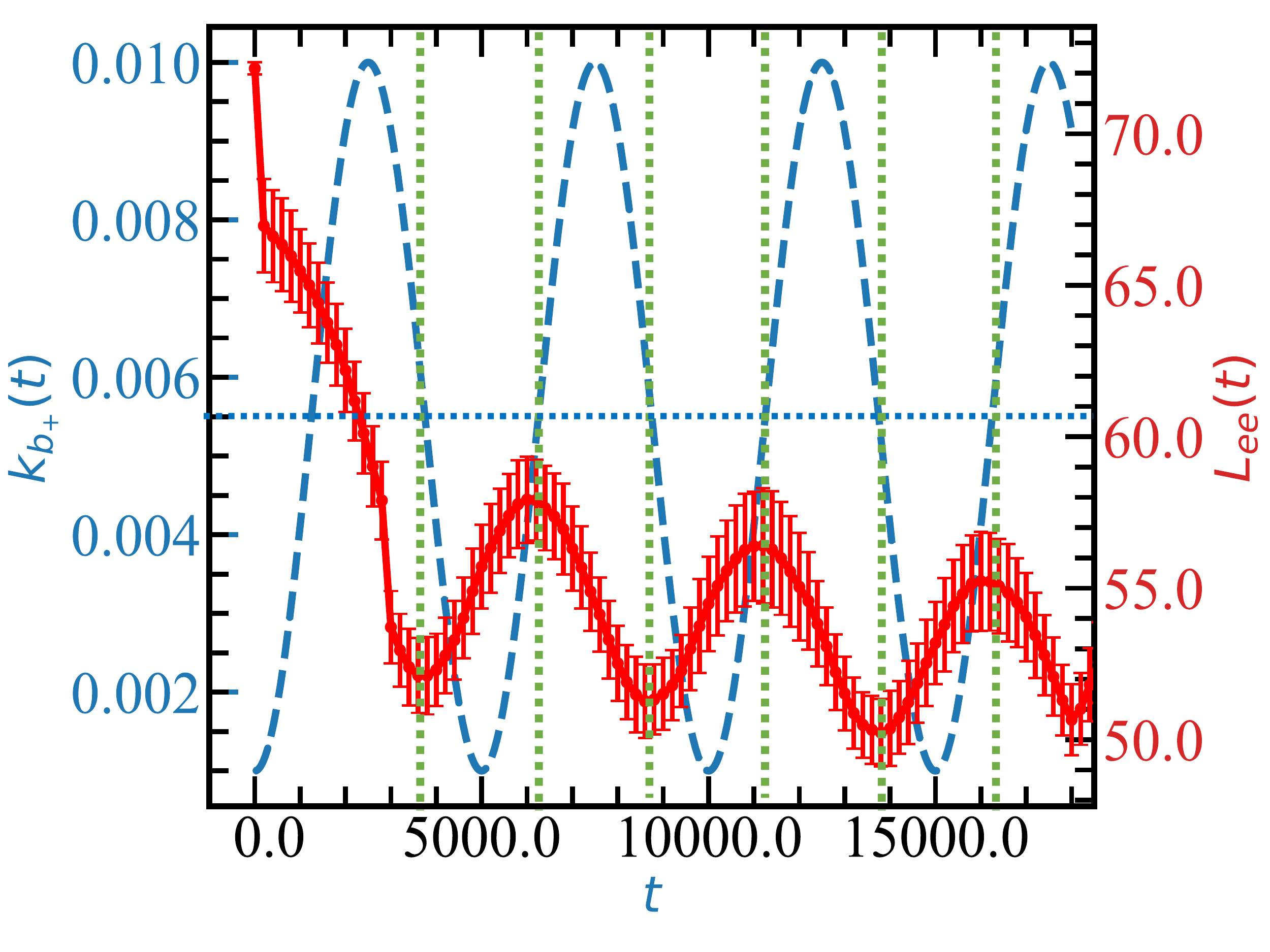}
}
         \caption{\label{fig:40f_T5000_eed_bm}
             (left ordinate, blue dashed line) Fluid-phase reaction rate coefficient $k_{b+}(t)$ versus time for oscillations with period $\tau_b=5000$, and (right ordinate, red line with error bars) the average end-to-end distance $L_{ee}$ of a filament in a 40-filament system. The horizontal blue dotted line marks $k_{b\pm}=5.5 \times 10^{-3}$ where $F_d=0$ and the motors are inactive, while the vertical green dotted lines mark every half period of the oscillation of $k_{b\pm}$. The results were obtained from averages over all filaments in the system and over 10 realizations of the dynamics.}
\end{center}
\end{figure}

Considering the fluid phase reaction model chosen in this work, and making use of Eq.~(\ref{eq:t-dependent-k}),
we show how an oscillating active gel-like state can be obtained by periodic variation of the effective rate coefficients. In particular, we take these rate coefficients to be given by
\begin{equation}\label{eq:bulk-rates}
k_{b\pm}(t)=\bar{k}_b \mp \Delta_b \cos \Omega t,
\end{equation}
where $\Omega =2 \pi /\tau_b$ with $\tau_b$ the period of the oscillation.
If we take $\bar{k}_{b}= 5.5 \times 10^{-3}$, the rate coefficient value for inactive motors, and $\Delta_b=4.5 \times 10^{-3}$, the rate coefficients will oscillate between those for forward-moving motors, ($k_{b+},k_{b-})= (10^{-2},10^{-3}$) at times $t=n \tau_b/2$ and backward-moving ($k_{b+},k_{b-})= (10^{-3},10^{-2}$) at times $t=n \tau_b$, for integer $n$. When such periodically-varying rate coefficients are used in the simulations of the active network, the mean end-to-end length of the filaments $L_{ee}(t)$ oscillates as shown in Fig.~\ref{fig:40f_T5000_eed_bm} (right ordinate). For reference, the figure also plots $k_{b+}(t)$ (left ordinate) given by Eq.~(\ref{eq:bulk-rates}).

To provide an analysis of the behavior of $L_{ee}(t)$ seen in this figure we consider the time scales of the chemical reaction rates and conformational changes that are important for the dynamics. The diffusiophoretic forces respond changes in the effective rate coefficients through the behavior of the $A$ and $B$ concentration fields that enter the motor catalytic reactions. In the network, which contains many motors, these fields may have a complex spatiotemporal structure. However, we can estimate these many-body effects using a mean-field description of the reaction kinetics,
\begin{eqnarray}
\frac{d}{dt} c_A(t) &=& -k_m n_c c_A(t) +k_m n_c c_B(t)\nonumber \\
&&-k_{b-}(t) c_A(t) + k_{b+}(t) c_B(t),
\end{eqnarray}
with a similar equation for $c_B$. Here $k_m$ is the rate coefficient for reactions on the catalytic spheres, and $n_c$ is the number density of catalytic spheres in the system. Using the condition $c_A + c_B =c_0$, where $c_0$ is the constant total concentration of reactive species, along with Eq.~(\ref{eq:bulk-rates}), we may write this equation as
\begin{eqnarray}
\frac{d}{dt} c_A(t) &=& -2(k_m n_c + \bar{k}_b) c_A(t) +(k_m n_c + \bar{k}_b)c_0\nonumber \\
&&-c_0 \Delta_b \; \cos \Omega t
\end{eqnarray}
whose solution is
\begin{eqnarray}
c_A(t)&=& e^{-t/\tau_{\rm ch}} c_A(0) \\
&&+\frac{c_0}{2}\Big(1- e^{-t/\tau_{\rm ch}}\Big)
+c_0 \frac{\Delta_b \tau_{\rm ch}^{-1}}{\tau_{\rm ch}^{-2} + \Omega^2}e^{-t/\tau_{\rm ch}} \nonumber \\
&&- c_0 \frac{\Delta_b}{\tau_{\rm ch}^{-2} + \Omega^2} \Big(\tau_{\rm ch}^{-1}\cos \Omega t +\Omega \sin \Omega t \Big),\nonumber
\end{eqnarray}
where the chemical relaxation time is $\tau_{\rm ch}=1/[2(k_m n_c + \bar{k}_b)]$. The rate coefficient $k_m$ can be written as $k_{m\pm}=k_\pm k_D/(k_+ +k_- +k_D)$, where $k_\pm=p_\pm R_c^2 (8 \pi k_BT/\mu)^{1/2}$ and $k_D=4 \pi D R_c$ with $R_c$ the effective radius of the catalytic sphere for interactions with the reactive species, and $\mu \approx m$ is the reduced mass of the colloid and solvent species. We neglect any screening by the noncatalytic spheres, the dimers and other filament beads. In the simulations we take $p_+=p_-$ so that $k_{m\pm}=k_m$. Using the system parameters given in the Appendix we find $\tau_{\rm ch} \approx 85$. The oscillation period is $\tau_b=5000$, and for times $\tau_{\rm ch} \ll t \ll \tau_b$ we have $c_A(t) \approx c_0 \big(\frac{1}{2}-  \tau_{\rm ch} \Delta_b \cos \Omega t \big)$. Using this result in Eq.~(\ref{eq:Fd}) gives rise to oscillatory diffusiophoretic forces,
\begin{equation}
F_d(t) = -2f_d k_+ \tau_{\rm ch} \Delta_b \cos \Omega t.
\end{equation}
This calculation shows that concentration fields, and thus the diffusiophoretic forces, are able to adapt very quickly to the much slower periodic variations of the effective rate coefficients, and points on the $k_{b+}(t)$ curve in the figure can be mapped onto the instantaneous values of the diffusiophoretic forces to a good approximation. These oscillatory diffusiophoretic forces, in turn, lead to oscillatory changes in the conformational structures of the active gel-like states.

Returning to a description of the results in Fig.~\ref{fig:40f_T5000_eed_bm}, after short initial transient period that depends on the initial preparation of the system, we see that $L_{ee}(t)$ oscillates with its minima and maxima corresponding to the points where $k_{b+}=5.5 \times 10^{-3}$ (indicated by vertical lines in the figure) and where $F_d$ changes sign. Consider the dynamics starting from the first minimum in the figure. As time increases over a half period, $k_{b+}< k_{b-}$ so that $F_d<0$, and in this time domain $L_{ee}(t)$ increases since the diffusiophoretic forces will tend to stretch the filaments. (Note that even as $k_{b+}(t)$ approaches $k_{b+}=5.5 \times 10^{-3}$ where $F_d=0$ the filament will continue to expand since its equilibrium state has an extended conformation.) However, before the filaments can fully extend the system enters next half cycle where $k_{b+}> k_{b-}$. Now $F_d>0$ and $L_{ee}(t)$ decreases. We can see from this mechanism that the minima of $L_{ee}(t)$ must lie at points where $F_d$ changes from positive to negative (forward to backward motors), and the maxima must lie at points where $F_d$ changes from negative to positive (backward to forward motors). Since the oscillations are simple harmonic functions, this implies that the extrema in $k_{b+}$ are phase shifted by a quarter of a period from those of $L_{ee}(t)$, as seen in the figure.~\cite{foot:complex}
\begin{figure}[htbp]
  \begin{center}
\resizebox{1.0\columnwidth}{!}{
     \includegraphics[height=6.5cm,width=10.5cm]{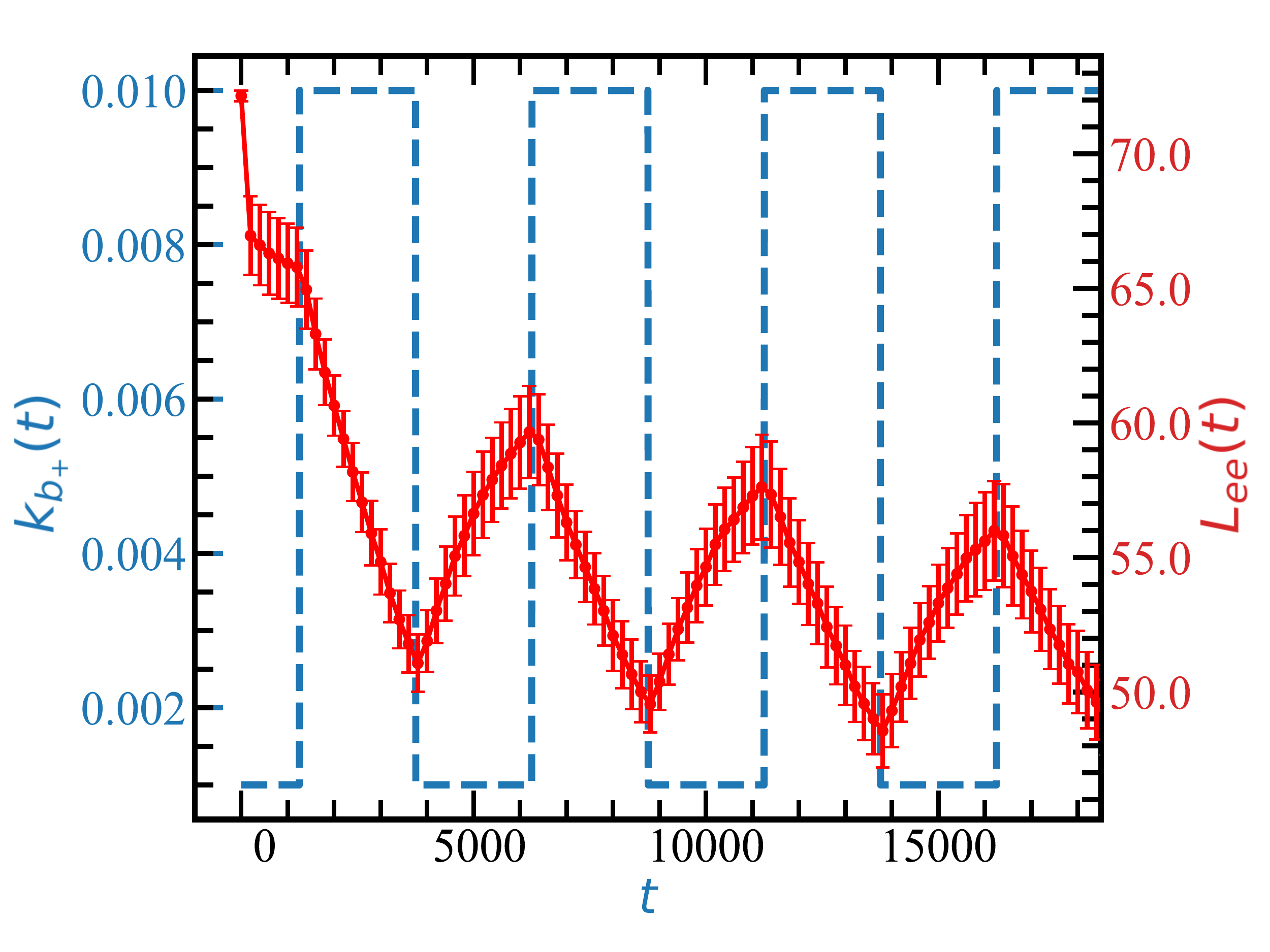}
}
         \caption{\label{fig:40f_1flength_t5k}
             Same as Fig.~\ref{fig:40f_T5000_eed_bm} except square-wave oscillatory variations of the fluid-phase reaction rate coefficients $k_{b\pm}(t)$ are applied.}
\end{center}
\end{figure}
The character of the dynamics depends on the form of the periodic oscillation. For example, if fluid-phase rate coefficients have a square-wave oscillatory form, one obtains the results shown in Fig.~\ref{fig:40f_1flength_t5k}. Because there are discontinuous switches in $k_{b\pm}$, the extrema again  correspond to these switch points, but there is no phase shift between $k_{b\pm}(t)$ and $L_{ee}(t)$ since between switches evolution is governed by the maximum diffusiophoretic forces for forward and backward motors corresponding to $(k_{b+},k_{b-})= (10^{-2},10^{-3})$ and $(10^{-3},10^{-2})$. By contrast, for harmonic oscillations, these forces change smoothly from zero at the switch points, reach their maximum absolute values, then return to zero values at the next switch. These factors are responsible for different forms of the oscillations in the two cases.

Lastly, we note that the homogeneity of the gel-like network also changes under periodic variation of the concentration constraints. The $NN$ radial distribution function $g(r)$ for a system with square-wave oscillation is shown in  Fig.~\ref{fig:grcu_40fila} (b). The curves were obtained averages over ten realizations of the dynamics and time averages over the time intervals $t=6000-7000$ (cyan diamonds and lines) and $t=8000-9000$ (purple triangles and lines). There is a prominent peak at $r=5.0$ in the $t=8000-9000$ data that corresponds to $NN$ clustering, while there is only very weak structural ordering for $t=6000-7000$. Thus, not only does the average end-to-end filament length change during the oscillation cycle, but so does the inhomogeneous structure of the filament system.

\section{Summary and Conclusion}\label{sec:conclusion}

Polymers that are able to respond to their environments or external stimuli have diverse applications.~\cite{roy2010,cohenstuart2010,Hilber2016,wei2017,brighenti2020} For example, the ability of responsive polymers to change their shapes forms the basis for devices that perform targeted drug delivery where drugs are encapsulated in the collapsed form of polymer nano-aggregates and released in their expanded forms. Similarly, switches and valves that open and close in microfluidic devices are constructed from polymers that respond to stimuli that effect such changes. On larger scales responsive polymers have been used to fabricate artificial muscles that are essential part of soft robotics applications.~\cite{wangGao2021} In these and other systems the response is often due to the fact that the state of the polymer depends on conditions such as pH, temperature, electric fields, etc.

By contrast, in this work we considered situations where chemically-active agents are incorporated in the polymers and used to control their conformational state. Thus, this work exploits the information gained in studies of the physics of soft active matter to obtain another perspective on the control of responsive polymer systems that makes use of some of the fundamental physical principles that underlie chemically-powered active motion.

The chemically-powered nanomotors embedded in polymeric filaments are not only the source of active conformational dynamics but also provide a means for the control of this activity. This control is achieved by making use of the basic microscopic reversibility of catalytic reactions on the motor surfaces along with fact that detailed balance is broken to promote active motion by chemical constraints on the system. Through such constraints the diffusiophoretic forces that the embedded motors exert on the filaments can be changed in a prescribed manner. In the model studied here the reversible reactions on the motors had equal rate coefficients so that the equilibrium constant was unity, and the motors were driven between forward and backward motion by changing the constraints. Often, diffusiophoretic motors studied in the laboratory have catalytic reactions that strongly favor production of product from fuel. In this case constraints can still be applied to effect the conformational changes discussed in this paper. For example, if the polymer is in a good solvent and its equilibrium state is an extended chain, then simply adjusting the reservoir to control the supply chemical fuel will induce conformational changes the chain length. Analogous arguments apply to equilibrium states where the polymer chains are collapsed. In systems containing many interacting filaments, different conformational states with stretched or partially collapsed filaments can be selectively accessed by such control of the constraints.

This link to active soft matter physics suggests ways to modify or improve the functionality of some of the devices mentioned above, as well as the construction new devices that make use of the properties of the embedded active elements. For instance, a main concern in biomedical applications is the biocompatibility of the chemical stimuli used to drive the conformational changes. There is a large literature on the use of enzymes to power nanomotor motion~\cite{MJAHMSS15,MHPS16,MHS16,ZGMS18,Petal18}. We noted earlier that dimer motors could be constructed with enzyme-coated catalytic spheres, so biocompatibility is much less difficult to achieve and enzymes can be chosen to respond to specific substrates. In this work we showed that by periodically varying the fluid phase effective rate constants that depend on pool species, which are fixed by reservoirs, cyclic expansion and contraction of the gel-like filament states was obtained. However, analogous to the use of the oscillating chemical concentrations in the Belousov-Zhabotinsky reaction in studies of active hydrogels, one may use fluid phase reactions that exhibit more complex oscillatory kinetics. Enzymatic chemical networks operating out of equilibrium are examples of systems that display sustained oscillations and bistability.~\cite{goldbeter1996}. Through the use of such oscillating enzymatic reactions to determine the nonequilibrium state of the system, autonomous control of the filament oscillations could also be achieved. We saw that additional features of the collective active filament states are present because of the embedded filament motors. The diffusiophoretic motors in bulk solution undergo active self assembly if they are forward moving but not when they move backward. Consequently, the many-filament systems with partially-collapsed filaments are highly inhomogeneous because the tendency of the motors to cluster but those with elongated filaments are more homogeneous and do not cluster.

The results presented here suggest other possibilities for constructing active filament systems. If permanent cross links among the filaments are included, active filament networks with two or three dimensional geometries can be constructed for materials science applications. In addition to providing another mechanism for constructing oscillating gel-like states, some features specific to embedded motor-filaments, such as the tendency of the embedded motors to form or prevent filament cluster formation, could be used to induce inhomogeneous strains in the network produce specific distortions of the system.

In the examples presented in this paper different chemical constraints were imposed by assuming that pool chemical species involved in fluid phase reactions could be varied, and a specific arrangement and choice of orientations of the embedded motors was made to induce elongation or contraction of the filaments. While the simulations in this paper use a given set of parameters and conditions to illustrate the properties of the active filament networks, many of the qualitative features of the phenomena we study do not rely an these specific parameter choices. For instance, to induce the chemomechanical coupling all that is needed is a diffusiophoretic force. Although its value is system specific and depends on rate coefficients, motor dimensions and fluid properties, the diffusiophoretic mechanism has been shown to operate in a wide variety of systems. Since the nonequilibrium states of these motors are determined by reservoir concentrations, the basic elements of control are present and can be implemented in various ways. Other choices of embedded motor configurations may give rise to different types collective filament states and this feature could be useful in the design of active materials and devices.

\appendix

\section{Computational Details}\label{sec:app}
Simulations of the dynamics of the active filament systems are carried out in a $75 \times 75 \times 75$ box with periodic boundaries. All filament beads have the same size with radius $\sigma_{\rm F}=1.5$ except those for the motor segments. The dimer motor segments have radius of $\sigma_{\rm C}=1.0$ and $\sigma_{\rm N}=2.0$. The beads are linked by stiff harmonic springs if their separation is less than the sum of the radii of the two beads. For the regular (non-motor) filament beads, $V_{\rm bond}^{FF}=\frac{1}{2}k_{s}(r-r_{\rm eq}^{FF})^2$, the equilibrium bond length is $r_{\rm eq}^{FF} = 1.0$. The dimer motor segments have $V_{\rm bond}^{CF}=\frac{1}{2}k_{sd}(r-r_{\rm eq}^{CF})^2$, $V_{\rm bond}^{NF}=\frac{1}{2}k_{sd}(r-r_{\rm eq}^{NF})^2$, $V_{\rm bond}^{CN}=\frac{1}{2}k_{sd}(r-r_{\rm eq}^{CN})^2$, where the equilibrium bond lengths are $r_{\rm eq}^{CF} = 1.75$, $r_{\rm eq}^{NF} = 2.75$, $r_{\rm eq}^{CN} = 3.5$, and the spring constants are $k_{\rm s} = 50$ and  $k_{\rm {sd}} = 100$. The bending stiffness of a filament is controlled by a three-body potential, $V_{bend}=\kappa_{b}[1-\cos  \theta ]$, with $\kappa_{b}=5.0$, and $\cos \theta =\hat{\bm{r}}_{i-1,i}\cdot \hat{\bm{r}}_{i,i+1}$, where $\hat{\bm{r}}_{i,j}=(\mathbf{r}_{i}-\mathbf{r}_{j})/\left | \mathbf{r}_{i}-\mathbf{r}_{j}\right |$. Also, it is necessary to have a strong enough repulsive LJ potential ($\epsilon_{D}=5.0$) for interactions between the monomers in the same filament to avoid the overlap between filament beads in strongly bent configurations.

The fluid phase contains $N=N_{A}+N_{B}$ particles of species $A$ and $B$. With the exception of the harmonic spring potentials discussed above and used to construct the filaments, all other intermolecular interactions take place through repulsive Lennard-Jones (LJ) potentials of the form
\begin{equation}
V_{\alpha \alpha'} =4\epsilon_{\alpha \alpha'} \Big[\Big(\frac{\sigma_{\alpha \alpha'}}{r_{i j}}\Big)^{12}-\Big(\frac{\sigma_{\alpha \alpha'}}{r_{i j}}\Big)^{6}+ \frac{1}{4} \Big]\theta(r_c-r_{i j}),
 \end{equation}
where $\theta(r)$ is the Heaviside function and the separation between a particle
$i$ of type $\alpha$ and a particle $j$ of type $\alpha'$ is $r_{i j} = |\bm{r}_i-\bm{r}_j|$. We let the symbols $\alpha,\alpha'=C,N$ denote the catalytic and noncatalytic monomers in the filament. The repulsive potential between two beads in different filaments has $\sigma_{\alpha \alpha}=2 \sigma_{\alpha}+\sigma$, $\sigma_{\alpha \alpha'}=\sigma_{\alpha' \alpha}= \sigma_{\alpha}+\sigma_{\alpha'}+\sigma$,
 $\sigma_{\alpha F}=\sigma_{F\alpha}= \sigma_{\alpha}+\sigma_{F}+\sigma$, with $\sigma=1.0$. Filament beads interact with other beads in neighboring filaments with strength $\epsilon_{FF}=1.0$. The interaction strengths of the repulsive interactions between motor beads and the filament are $\epsilon_{\alpha \alpha}=\epsilon_{\alpha \alpha'}= \epsilon_{\alpha'\alpha}=\epsilon_{\alpha F}=\epsilon_{F\alpha}= 1.0$. The $A$ and $B$ fluid particles have identical effective radii $\sigma_{A}=\sigma_{B}=0.25$ , and energy parameters $\epsilon _{AC}=\epsilon _{BC}=\epsilon _{NA}=1.0$, $\epsilon _{NB}=0.1$ and $\epsilon_{AF}=\epsilon_{BF}=0.1$ for their interactions with the polymer beads.

All solvent species have the same mass $m$, whereas masses of the motor and the filament beads are chosen to be $m_{\alpha} = (d_{\alpha}/d_{S})^{3}$ m so that they have the same mass density as a solvent particle. The average solvent density is $n_{0}=N/L^{3} \sim 9$.

The hybrid multiparticle collision dynamics-molecular dynamics simulation method consists of free streaming and collision steps.~\cite{Malevanets_Kapral_99,Malevanets_Kapral_00} In the streaming step, the dynamics of all the species is governed by molecular dynamics and propagated by Newton's equations of motion. In this step there is no net force among solvent particles. Instead, the interactions among the solvent particles are described by multiparticle collisions dynamics.\cite{kapral:08,gompper:2009}
In the collision step, at discrete times $\tau$, the system is divided into cubic cells $\xi$ with size $a_{0}=1$. The rotation operators $\hat{\omega}_{\alpha}=\pi/2$ are assigned to each cell from some set of rotation operators.  The post-collision velocity $\mathbf{v}_{i}(t+\tau)$ of each particle $i$ within the same cell can be obtained according to the rotation rule
$\mathbf{v}_{i}(t+\tau)=\mathbf{v}_{cm}(t)+\hat{\omega}_{\alpha}(\mathbf{v}_{i}(t)
-\mathbf{v}_{cm}(t))$, where the center-of-mass velocity $\mathbf{v}_{cm}$ of each cell $\xi$ is calculated from $\mathbf{v}_{cm}= \sum_{j=1}^{N_{c}}\mathbf{v} _j/N_{c}$ where $N_{c}$ is the total number of particles in the cell. Grid shifting is employed to ensure Galilean invariance.~\cite{Ihle_Kroll_01}

The fluid phase reactions as assumed to take place under nonequilibrium conditions by a mechanism
$B +P_1 \rightleftharpoons  A +P_2$, where the ``pool" chemical species $P_1$ and $P_2$
are assumed to be in excess and the values of their concentrations are incorporated in the
rate coefficients $k_{b\pm}$. The reactive version of multiparticle collision dynamics is used to carry out the reactive dynamics in the fluid phase~\cite{Rholf2008}.

The molecular dynamics time step is $\Delta t_{MD}=0.001$ for the solution of Newton's equations using the velocity-Verlet algorithm. The multiparticle collision time is $\Delta t_{MPC}=0.1$. The system temperature is $k_{B}T=0.2$. The viscosity of the fluid is $\eta=1.282$ and the fluid-particle self-diffusion coefficients are given by $D_{A}=D_{B}=D_{0}=0.118$.

In our simulations, all quantities are reported in dimensionless units based on energy in units of $\epsilon$, mass in units of $m$ and distance in units of $\sigma$.

\section*{Conflicts of interest}
There are no conflicts of interest to declare.

\section*{Acknowledgements}
\label{sec:acknowledge}
We thank Mu-Jie Huang of the University of Toronto for useful discussions.
This work was supported by the National Natural Science Foundation of China (12004086,11974094) and the Natural Sciences and Engineering Research Council of Canada and Compute Canada.

\bibliography{references}

\begin{thebibliography}{81}
\bibitem [{foo({\natexlab{a}})}]{foot:metastable}%
  \BibitemOpen
  \href@noop {} {\bibinfo {title} {For nearly linear filaments, such as those
  in the transient period, the diffusiophoretic forces are directd predominaly
  along the chain and chemomechanical coupling is weak; thus, a transient
  period exists since sufficiently large fluctuations are needed to induce
  conformational transitions that lead to the final steady or equilibrium
  states.}} ({\natexlab{a}})\BibitemShut {NoStop}%
\bibitem [{foo({\natexlab{b}})}]{foot:complex}%
  \BibitemOpen
  \href@noop {} {\bibinfo {title} {For more complex oscillatory fluid phase
  reactions simple harmonic oscillations occur close to a hopf bifurcation,
  while farther from this point the oscillations may adopt relaxational or more
  complicated structures.}} ({\natexlab{b}})\BibitemShut {NoStop}%
\end{thebibliography}

\end{document}